# Some New Methodologies for Image Hiding using Steganographic Techniques

Rajesh Kumar Tiwari and Gadadhar Sahoo

*Abstract*—Security and memory management are the major demands for electronics devices like ipods, cell phones, pmps, iphones and digital cameras. In this paper, we have suggested a high level of security mechanism by considering the concept of steganography along with the principle of cryptography. Four different methods that can save a considerable amount of memory space have been discussed. Based on these methods, we have constructed secured stego image creator and secured multi image viewer in Microsoft platform so as to provide high level of security and using less memory space for storage of image files in the above said electronic devices.

## I. INTRODUCTION

The growing use of Internet need to store, send and receive personal information in a secured manner. For this, we may adopt an approach that can transfer the data into different forms so that their resultant data can be understood if it can be returned back into its original form. This technique is known as encryption. However, a major disadvantage of this method is that the existence of data is not hidden. If someone gives enough time then the unreadable encrypted data may be converted into
its original form.

A solution to this problem has already been achieved by using a technique named with the Greek word "steganography" by which we mean 'writing in hiding'. The main purpose of steganography is to hide data in a cover media so that other cannot notice it [10]. The characteristics of the cover media depends on the amount of data that can be hidden, the perceptibility of the message and its robustness 4]-[6], [8], [10].

Publishing and broadcasting fields also require an alternative solution for hiding information. Unauthorized copying is hot issue in the area like music, film, book and software. To overcome this problem some invisible information can be embedded in the digital media in such a way that no one can easily extract it [1], [2], [4]. Analogously, software industries have taken advantage of another form of steganography, called watermarking, which is used to establish ownership, identification, and provenance [3], [7].

Gadadhar Sahoo is with the Department of Computer Science & Engineering, Birla Institute of Technology, Mesra, Ranchi, India (phone: 91-651-2276185; fax: 91-651-2275401; email: gsahoo@bitmesra.ac.in).

Rajesh Kumar Tiwari is with the Department of Computer Science & Engineering, R.V.S. College of Engg. & Tech. Jamshedpur, India (e-mail: rajeshkrtiwari@yahoo.com).

## II. RELATED WORKS

The most suitable cover media for Steganography is image on which numerous methods have been designed. The main reason is the large redundant space and the possibility of hiding information in the image without attracting attention to human visual system. In this respect, a number of techniques have been developed [1], [7] using features like

- Substitution
- Masking and Filtering
- Transform Technique

The method of substitution generally does not increase the size of the file. Depending on the size of the hidden image, it can eventually cause a noticeable change from the unmodified version of the image [4], [6]. Least Significant Bit (LSB) insertion technique is an approach for embedding information in a cover image. In this case, every least significant bit of some or all of the bytes inside an image is changed to a bit of the sink image. When using a 24-bit image, one bit of each of the primary color components can be used for the above purpose. The masking and filtering techniques starts with the analysis of the image. Next, we find the significant areas, where the hidden message will be more integrated to cover the image and lastly we embed the data in that particular area. In addition to the above two techniques for message hiding, transform techniques has also been employed in embedding the message by modulating coefficients in a transform domain. As an example, we may mention here that Discrete Cosine Transform works by using quantization on the least important parts of the image in respect to the human visual capabilities.

The poor quality of recover image is the main drawback with the proposed LSB algorithm by Anderson and Petitcolas [11]. Raja et.al. [12] have proposed the least significant bit-embedding algorithm where cover image should always at least eight times larger than the sink image. As a result of which it can be pointed out here that a small image may not keep a large image. Masking Filtering and Transform techniques also have same limitations.

In order to overcome the above drawbacks, we propose here four methods that take care of storing one image file in another image and maintain more security and use comparatively less space. These methods are discussed in the next section. The implementation of the methods is given in section 4 followed by conclusion.

### III. PROPOSED METHODS

In this section, we propose four methods for image hiding where we store one image file, called sink image in another image file called as container image. The primary objective is to use steganography techniques so as to provide more security and simultaneously using less storage. Digital images are commonly of two types i) 8 bit images and ii) 24 bit images. Here, we virtually partition both container and sink images into two parts namely, structural part and data part. In structural part, we keep all structural information of the image file like file header information, coloring palette etc. In the data part the actual image pixel values are being considered. The four different parts for container and sink images will be

- Structural Part of Container Image
- Data Part of Container Image
- Structural Part of Sink Images
- Data Part of Sink Images.

Our main intention is also not to change the visual properties of both images by which the structural part of container image will be remained same for entire hiding process. We next explain the partitioning of bmp image here just for illustration purpose.

A BMP file is generally composed of 3 or 4 parts. If the file contains 3 parts, the structure of corresponding BMP file can be shown as in Fig.1a. Otherwise the same file can take the form as in Fig.1b. We consider Header, Info Header, and Optional Palette in structural part, whereas in the data part we consider only the image data.

For storing both structural part and data part of the corresponding sink image file in a container image file, we propose here the following four different methodologies.

### A. Methodology-I (Addition Method)

In this method we are storing both parts of the sink image at the end of the data part of the container image with a stego key, finally we get the stego image, which contains container image and sink image together. In view of the fact that the structural part of container image is unchanged, the visual appearance of both container image and new stego-image will be remain same. The advantage of this method is solely due to its simplicity, which can easily be seen from Fig. 2.

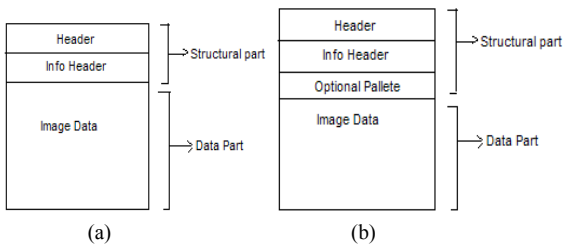

Fig. 1. (a) BMP Image Formats with 3 parts. (b) BMP Image Format with 4 parts.

The pseudo code for the above methodology is given below.

/* Pseudo Code for Methodology-I */

Addition Method
{
Array container_image []=LoadImage("Container Image", Length);

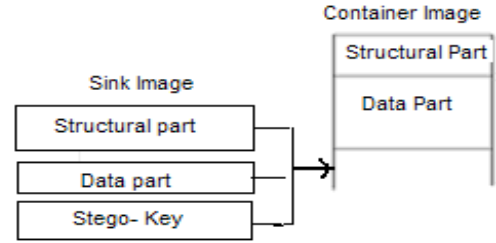

Fig. 2. Storing data at the end of the data part of container image.

Array sink_image [] = LoadImage ("Sink Image", Length);

Array stego_image;

String stego_key;

Integer cont_data_part,cont_str_part,snk_data_part,snk_str_part;

Integer i, j ,k;

Input cont_data_part, cont_str_part,snk_data_part,snk_str_part;

For i=0 to cont_str_part
    Stego_image [i] = container_image [i];
End for;

For i= cont_str_part to cont_data_part
    Stego_image [i] = container_image [i];
End for;

For j=0 to Length (stego_key)
    Stego_image [j] = stego_key [i];
End for;

For k=0 to snk_str_part
    Stego_image [k] = sink_image [k];
End for;

For k= snk_str_part to snk_data_part
    Stego_image [k] = sink_image [k];
End for;

SaveImage (stego_image [])

RemoveImage(container_image []);

RenameImage (stego_image, container_image);
}

### B. Methodology-II (LSB Adjustment Method)

In this method, we break the storing process into two steps,
- Storing structural part, and
- Storing data part in different positions.

Since structural part of an image takes less space comparing to the data part, the structural part of the sink image may be stored in the data part of the container image by using the most common least significant bit technique. Consequently we can say a number of bits corresponding to a sink image. That means if M bits represents the original structured part of the sink image, we need only N (< M) bits by the use of LSB techniques. It may be mentioned here that when one wants to store an adequate number, k, of sink images one can effectively save $\sum M_i - N_i$ bits (where $1 \leq i \leq K$) or which may be quite large in size. Further, for security point of view, as we are inserting the stego-key before and after the structural part of each sink image it gives an additional layer of protection (Fig. 3).

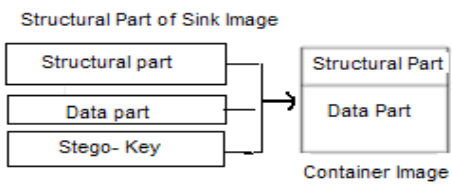

Fig. 3. Storing structural part of sink image.

One of the most important considerations while designing technique to be used for image hiding is that it should perform its operation without raising any suspicion of the eavesdropper. Most steganographic technique implicitly employs limitations of the human hearing systems or human visual system to embed the data. The requirement of larger space for the data part of the sink image may draw our attention to store this part at the end of the data part of the stego image. This leads to get advantages: first the visual changes will be less and second, a small container file may keep a large sink file (Fig. 4).

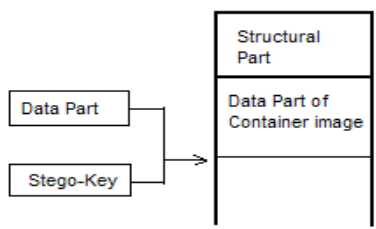

Fig. 4. Storing data part of sink image.

For the above purpose, we adopt the following principle to store more sink files in a single container file by finding the appropriate point (address) on it.

Let us consider L bits is the total space of the data part of the container image, n is the number of sink image to be stored, K bits is the space of the each structural part of sink image, and $Z(1<=Z<=L)$ be the position of the storing place in the container image. Then, we define

N x P < M       (1)

and

Z = N (X + Y) + C       (2)

where, x is the structural part data bits of sink image, y is the data part data bits of sink image and c ($1<=c<=k$) is the individual sink image. For example, if we are having six sink images and the first bit of the structural part of the first sink image will give the destination place, then for N = 6, X = 1, Y = 0 and C = 1, we get Z = 7. It means that the corresponding information will be stored at the least significant bit of the 7[th] byte of the container image. Similarly, for the second bit of the second sink image we may have Z=14 and will be stored at 14[th] byte of the container image. We can apply the same formula for the data part also. The required pseudo code is given below.

/* Pseudo Code for Methodology-II */

LSB Method
{
   Array cont_ image [ ] = Loadimage ("Container Image", Length);

   Array sink_image [ ] = LoaderImage ("Sink Image", Length);

   Array stego_image [ ];

   Boolean snk_str_part_bit [ ];

   String stego_key;

   Integer cont_data_part, cont_str_part, snk_data_part, snk_str_part;

   Integer i, j, k, m, byte_value;

   Input cont_data_part, cont_str_part, snk_data_part, snk_str_part;

   m=1;

/ * storing structure part of sink image */

   For i = 1 to snk_str_part
       byte_value = sink_image[i];

     For j = 0 to 7
       snk_str_part_bit[m] = MOD (byte_value, 2);
       byte_value = byte_value / 2;

      End For;

   End For;

   k = cont_data_part;

   For i = 1 to LENGTH (snk_str_part_bit [ ])

   If (snk_str_part_bit[i] <> MOD (cont_image[k], 2) Then

     If (snk_str_part_bit[i] = 0) then
       cont_image[k] = Asc (cont_image[k]) – 1;
     Else
       cont_image[k] = Asc (cont_image[k]) + 1;
     End if;

   End if;

  End if;
     k=k+1;
  End For;

/ * creating stego_image */

```
    For i = 0  to cont_str_part
        stego_image[i] = container_image[i];
    End For;

    For i = cont_str_part to cont_data_part
        stego_image[i] = container_image[i];
    End For;
        For j = 0 to Length (stego_key)
    stego_image[j] = stego_key[j];
        End For;
    For k = snk_str_part to snk_data_part
        stego_image[k] = sink_image[k];
    End For;
    SaveImage (stego_image [ ] );
    RemoveImage (container_image [ ]);
    RenameImage(stego_image, container_image);
}
```

### C. Methodology-III (Structural Method)

In this approach we consider images having similar structural parts for the reason that instead of saving structural part of an image again, we can use the structural part of the previously stored images. And, the data part of sink image can be stored in the data part of the stego- image with the help of least significant bit method. By using this method, maximum space can be saved (Fig.5).

/* Pseudo Code for Methodology-III */

Structural Method

{

Array cont_image []=LoadImage("Container Image", Length);

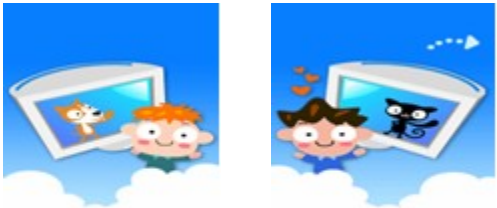

Fig. 5. Images having similar structural parts.

Array sink_image [] = LoadImage ("Sink Image", Length);

Array stego_image;

Boolean snk_data_part_bit[ ];

String stego_key, Flag;

Integer cont_data_part,cont_str_part,snk_data_part,snk_str_part;

Integer i, j ,k, m, byte_value;

Input cont_data_part, cont_str_part,snk_data_part,snk_str_part, Flag;
m=1;

/* Checking Similar Structural Portion */

For i=0 to snk_str_part

  If ( cont_image [i] = sink_image [i] ) then
      Continue;
  Else
      cont_image [i]= Flag;
End for;

/* Storing Data part of Sink Image * /

For i= 1 to snk_data_part
    Byte_value= sink_image [i];

  For j=0 to 7

    Snk_data_part_bit [m] = MOD (byte_value, 2);

    Byte_value=byte_value/2;
  End for;

End for;

For j=0 to Length ( stego_key)
  Stego_image [j] = stego_key [i];
End for;

K= cont_data_part;

For I =1 to Length ( snk_data_part_bit[])
   If ( snk_data_part_bit[i]=0) then
      Cont_image[k] = Asc( cont_image[k] -1);
   Else
      Cont_image[k] = Asc( cont_image[k] +1);
 End for;

/* Creating stego_image */

For k=0 to snk_str_part
   Stego_image [k] = sink_image [k];
End for;

For k= snk_str_part to snk_data_part
   Stego_image [k] = sink_image [k];
End for;

SaveImage(stego_image[])

RemoveImage(container_image []);

RenameImage(stego_image, container_image);
}

### D. Methodology-IV (Data Method)

In this technique we have given more emphasis on the data part of sink images. First, we check the similarities between data part of sink image and the data part of the stego-image. Wherever we find the similar data portions, we place one flag there and repeat this step till the similar data portion is over. It is then followed by either of the approach LSB adjustment (Methodology-II) or structural method (Methodology–III) that we have already discussed in this paper. If we take Fig. 5 & 6, as an example, where we may get lot of similar portions which can be avoided for restoring in stego image. This particular methodology may give more fruitful result as more number of sink images can be stored in stego image. By this technique for the above said example we can save at least twenty five to thirty five percent on memory space for the sink image. In this method also the security aspect can be achieved by defining separate stego key.

```
/* Pseudo Code for Methodology-IV*/
Data Method
{
    Array cont_image[] = LoadImage("Container Image", Length);
    Array sink_image [] = LoadImage ("Sink Image", Length);
```

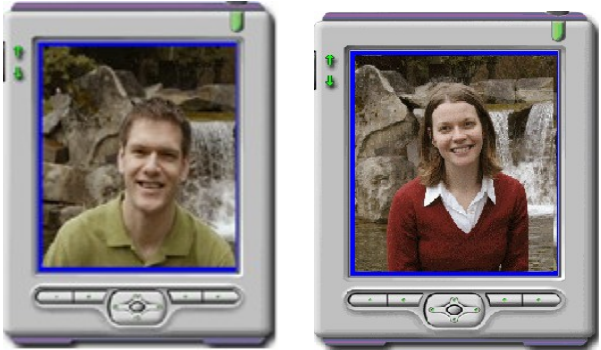

Fig. 6. Images having similar portion in their data parts.

```
    Array stego_image [];
    Boolean snk_data_part_bit [];
    String stego_key, Flag;
    Integer cont_data_part, cont_str_part, snk_data_part, snk_str_part;
    Integer i, j, k, m, byte_value;
    Input cont_data_part, cont_str_part, snk_data_part, snk-str_part, Flag;
        m=l;
        /* checking similar structure part */
        For i = 1 to snk_str_part
         If (cont_image[i] = sink_image[i]) then
            Continue;
         Else
            Cont_image[i] = Flag;
         End if;
        End For;

        /* checking and storing dissimilar Data Portion */
        For i = snk_str_part to snk_data_part
         For j = cont_str_part to cont_data_part
            If  (cont_image[j] = sink_image[i]) then
                Continue;
            Else
                Cont_image[i] = Flag;
            End if
         End For;
        End For;
/* creating stego_image */
For k=0 to snk_str_part
    Stego_image [k] = sink_image [k];
End for;
For k= snk_str_part to snk_data_part
    Stego_image [k] = sink_image [k];
End for;

SaveImage(stego_image[]);
RemoveImage(container_image []);
RenameImage(stego_image, container_image);
}
```

## IV. SSIC AND SMIV AND THEIR IMPLEMENTATION

With the help of the above defined methods, we can construct Secured Stego Image Creator (SSIC) and Secured Multi Image Viewer (SMIV), to provide more security and use less memory space for storing image files in electronic devices.

Security and space management are the major demand for the electronics devices like ipods, cell phones, pmp, iphones and digital camera. So, we can take a single stego image file for storing all required sink image files. As a result of which, not only a comparatively less space to be used but also more security can be achieved. Therefore, we can mention here that this concept will definitely be useful not only in the above said devices but also in the field of embedded system and even in our personal computer where each separate image takes separate space and maintains less security.

Here, we have implemented the above SSIC concept in Microsoft platform (fig. 7a). It may also be implemented in other programming platform like JAVA & C. In the Initial step, we check the structural part of the sink image, if it is matched with the previously stored structural parts then we apply methodology - III; else based on the other limitations we may choose methodology-I or methodology-II. In the second step, we store the data part of the sink image by using the methodology-IV. The Flowchart showing the details of this process is given in Fig. 8. It has been seen from our experiment on few images of fig.9 depicting some images of similar structural and data portions that for an average condition the use of SSIC can save fourteen to eighteen percent of memory space (fig.10). Furthermore, in contrast to the principle of cryptography, in this case the security level becomes high as we use separate stego key for each sink image file. It is worth mentioning here that as we are not doing any major changes on container and sink image pixel values, the visual properties of all images will be remain same.

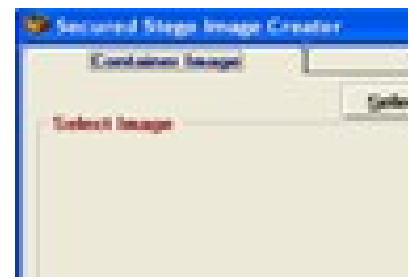

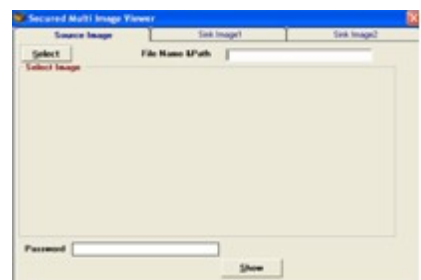

Fig. 7. (a) Secured Stegoo Image Creator (b) Secured Multi Image Viewer.

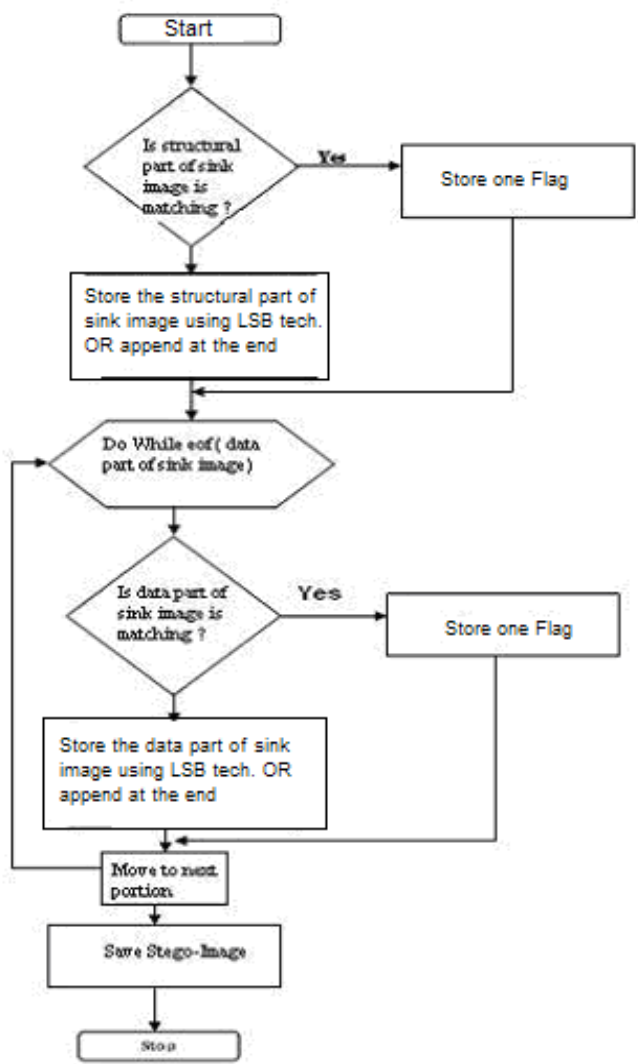

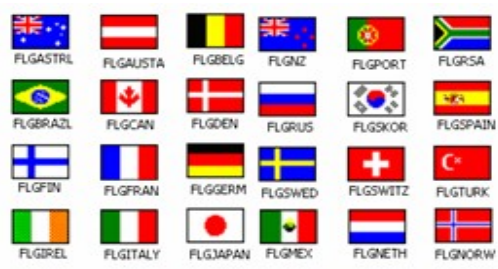

Fig. 9. Images having similar structural and data portions.

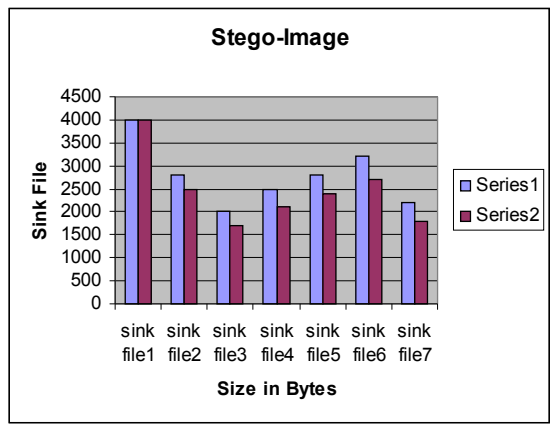

Fig. 10. Graph showing relation between the size before and after uses of SSIC

Secured multi image viewer (fig.7b) is the reciprocal part of the SSIC. The decoder program first loads the Stego image. Then, after giving correct stego-key, the required processing starts. In first step of processing, the structural part is retrieved and then the data part is appended. Finally the container and sink images are displayed in different grids, so that we can view all sink images stored with a particular stego key at the same time.

## V. CONCLUSION

In this paper, we have presented four different methodologies for image hiding based on steganography. It can be concluded here that the high degree of redundancy that generally found in a digital representation of multimedia content can give an opportunity to reduce the memory space. Instead of using different locations for storing different image files, it is possible and sufficient to use a single image file for the same purpose and with high level of security. The given concepts may be very much useful for the devices like ipods, cell phones, pmps, iphones digital cameras, and even in the personal computers where each separate image file take separate memory space and maintain less security. In place of sink images, we can even apply them to store messages of large sizes.